\documentclass[print]{aa} 
\usepackage{graphicx}
\usepackage{subfig}
\usepackage{txfonts}
\usepackage[flushleft]{threeparttable}
\usepackage{mathtools}
\usepackage{multirow}
\usepackage{colortbl}
\usepackage{comment}
\usepackage{stfloats}
\usepackage{float}
\usepackage{afterpage}
\usepackage{hyperref}  
\hypersetup{colorlinks=true,linkcolor=[rgb]{0.1,0.4,1.},citecolor=[rgb]{0.1,0.4,1.},filecolor=[rgb]{0.7,0.2,0.2},urlcolor=[rgb]{0.1,0.4,1.}}

\usepackage{color}
\definecolor{blue}{rgb}{0., 0., 1}
\definecolor{lightblue}{rgb}{0.1,0.4,1.}

\usepackage{soul}
\usepackage{multicol}

\begin{document} 
\title{An intriguing coincidence between the majority of vast polar structure dwarfs and a recent major merger at the M31 position} 

\titlerunning{Coincidence in 6D between the LMC, VPOS dwarfs and M31 merger}
\authorrunning{I. Akib et al.}

\author{Istiak Akib\inst{\ref{ins1}} \fnmsep\thanks{E-mail: \href{mailto:istiak-hossain.akib@obspm.fr}{istiak-hossain.akib@obspm.fr}}
\and François Hammer\inst{\ref{ins1}}
\and  Yanbin Yang\inst{\ref{ins1}}
\and Marcel S. Pawlowski\inst{\ref{ins2}}
\and Jianling Wang\inst{\ref{ins1}}}

\institute{
LIRA, Observatoire de Paris, Universite PSL, CNRS, Place Jules Janssen 92195, Meudon, France \label{ins1}
\and
Leibniz Institute for Astrophysics, An der Sternwarte 16, 14482 Potsdam, Germany \label{ins2}}

\date{Received 5 September 2024/ Accepted 20 December 2024}

\abstract{
A significant part of the Milky Way (MW) dwarf galaxies orbit within a Vast POlar Structure (VPOS), which is perpendicular to the Galactic disc and whose origin has not yet been identified. It includes the Large Magellanic Cloud (LMC) and its six dynamically associated dwarf galaxies. Andromeda Galaxy (M31) experienced a major merger two to three billion years ago, and its accurate modelling predicts that an associated tidal tail is pointing towards the Galaxy. Here, we tested a possible association between M31 tidal tail particles and MW dwarf galaxies, focusing first on the LMC and its associated dwarfs since they are less affected by ram pressure. We traced back these dwarf galaxy orbits by one billion years and calculated their association with the tidal tail particles in the 6D phase space, based on their proper motion from \textit{Gaia} DR3. We find that for low-mass MW models (total mass less than 5 $\times 10^{11} M_{\sun}$), the separation in the 6D space can be less than 1$\sigma$ for most of the M31 modelling, albeit with a significant degree of freedom due to the still unknown proper motion of M31. We further discover that many other dwarfs could also be associated with the M31 tidal tails if their motions had been radially slowed, as expected from the ram pressure exerted by the MW corona. This intriguing coincidence could explain the origin of the VPOS, which resulted from a matter exchange between M31 and MW.  
}


\maketitle

\section{Introduction}
 The observational properties of the M31 and its outskirts are well reproduced by a recent major merger \citep{Hammer2018, DSouza2018, Bhattacharya2023,Tsakonas2024}. According to \citet{Hammer2018} simulations, the merger started 7-10 Gyr ago and ended 2-3 Gyr ago. These simulations are able to reproduce the age-velocity dispersion of the M31 stars in the disc, the bar, the bulge, the 10 kpc ring, the outer structures of the disc (shells and clumps), the giant stellar stream (GSS), and the slope of the halo profile. The simulation suggests that the GSS is made of the stellar particles ejected in a tidal tail and recaptured by the M31 post-merger galaxy. In this scenario, the GSS results from the superposition of stellar loops orientated towards the observer. This implies that the Milky Way (MW) is currently located within the orbital plane of the M31 merger. This is also supported by observations of the M31 dwarf galaxies, half of which contribute to a satellite galaxy plane that is aligned edge-on towards the MW to within 1 degree \citep{Ibata2013}. In the M31 merger models, this satellite plane coincides with the orbital plane of the merging galaxies \citep{Hammer2013}, which favours the possibility that associated tidal tails extend to the MW. These tidal tails may be carrying a considerable amount of baryonic material, stars, or gas towards the MW (approximately a few times $10^{10} M_{\sun}$).

The MW is surrounded by about 50 known dwarf galaxies \citep{Li2021}, of which $\sim$20 seem to orbit in a plane perpendicular to the Galactic disc. This preferentially orientated structure of dwarf galaxies has been called the Vast POlar Structure \citep[VPOS;][]{Pawlowski2019,Pawlowski2021} and was first identified almost half a century ago \citep{Lynden-Bell1976}. It is not only identified from the dwarf locations but also from their proper motions, which indicate that they co-orbit along the VPOS \citep{Pawlowski2019, Pawlowski2021}. However, there is no well-accepted explanation for its existence. Orbital energies indicate that these dwarfs arrived at the MW within the last 3 Gyr \citep{Hammer2023}, and orbital analysis also suggests that the VPOS is a young or recent feature \citep{Taibi2024}. 

Dwarf galaxies can form in the tidal tails emerging from galaxy mergers through the collapse of the stellar and gas contents \citep{Wetzstein2007}. \cite{Pawlowski2011} showed that tidal debris from another galaxy can create both prograde and retrograde motion of the tidal dwarf galaxies (TDGs), as seen in the VPOS. \citet{Fouquet2012} proposed that the VPOS dwarfs formed in a tidal tail associated with the M31 merger. They found that the velocity of the tidal tail is consistent with the Large Magellanic Cloud (LMC) motion and were able to reproduce the VPOS geometry. \citet{Erkal2020} have shown that six other VPOS dwarfs have orbital motions linked to the LMC, which is by far the heaviest member of the VPOS. The M31 merger time frame is also consistent with the predicted infall time for the dwarfs. However, the analysis of \cite{Fouquet2012} was limited by the unknown proper motions for most of the dwarfs at a time well before the \textit{Gaia} era. We have a much better estimate of the proper motion of dwarfs \citep{Li2021, Battaglia2022} with \textit{Gaia}, as well as better estimates of the MW potential from rotation curves, which hints at a relatively low mass for the MW \citep{Jiao2023, Ou2023}. This enables us to compare the 3D position and velocity of the VPOS dwarfs with the infalling tidal tail particles.  

In this study we aim to test whether the VPOS dwarfs could originate from the M31 merger tidal tail. In Sect. \ref{sec: dwarf traceback} we trace back the VPOS dwarfs to investigate their origin and the similarity of their orbits. In Sect. \ref{sec:6D association} we calculate the infall of the tidal tail particles from the five M31 merger models reported by \cite{Hammer2018} and compare them in the 6D space with the VPOS dwarfs originating from a similar direction. In Sect. \ref{sec: discussion} we discuss whether a massive dwarf like the LMC could form in such a tidal tail and why there are four VPOS dwarfs among the sixteen that we could not associate with the tidal tail from the M31 merger. Section \ref{sec: conclusion} summarises our results. Throughout this paper, we use two rectangular coordinate systems: the current Andromeda-centric coordinates (in capital letters), where the current MW is on the z-axis, and the Galactocentric coordinates (in small letters).

\section{Tracing back the VPOS dwarfs}
\label{sec: dwarf traceback}

\begin{table*}[t]
\centering
\caption{\label{Table pot model}Parameters for the MW potential models. The CGM and DM profiles are shown in Appendix \ref{sec: MW Dark Matter Models}.}
\centering
\begin{tabular}{ccccccccccc}
\hline
\hline
 {MW}& {$M_{tot}$} & $R_{200}$&\multicolumn{2}{c}{Baryon} & \multicolumn{4}{c}{Dark Matter} \\

  Model&{\tiny $10^{11} \mathrm{M_{\odot}}$}& kpc & Model & $M_{bary}$ ($10^{11} \mathrm{M_{\odot}}$) & Model & $M_{DM}$ ($10^{11} \mathrm{M_{\odot}}$) & $ \rho_0$$\mathrm{{(M_{\odot}pc^{-3})}}$ & $r_0$ (kpc) & n \\
\hline
 A & 2.06 & 109 & Bary1     & 0.616  & \multirow{4}{*}{Einasto} & 1.44 & 0.01992 & 11.41  & 0.43\\
 B & 2.42 & 109 & Bary1+CGM & 0.976  &  & 1.44 & 0.01992 & 11.41  & 0.43\\
 C & 4.92 & 153 & Bary2     & 0.9055 &  & 4.01 & 1.5083 & 0.0574 & 3\\ 
 D & 6.89 & 175 & Bary2     & 0.9055 &  & 5.98 & 8.0064 & 0.0036 & 4\\ 
 \hline
 E & 8.14 & 186 & Bary2     & 0.9055 & NFW     & 7.25  & 0.0106 & 14.8 &\\ 
\hline
\end{tabular}
\label{tab:MW models}
\end{table*}

To trace back the VPOS dwarfs, we used five MW potential models that have been used to fit the \textit{Gaia} DR2 or DR3 rotation curve. Table \ref{tab:MW models} summarises the MW models. The models are ordered from A to E with increasing mass ($M_{tot} = M_{bary} + M_{DM}$) values. $R_{200}$ is the radius of a sphere within which the average dark matter (DM) density is equal to 200 times the critical density of the Universe $ \rho_{cr} = 1.34 \times 10^{-7} \mathrm{M_{\odot} pc}^{-3}$ \citep{Hinshaw2013}, and $M_{DM}$ is the DM mass within $R_{200}$. Model A is from \cite{Jiao2023} and has been designed to reproduce the rotation curve using the \textit{Gaia} DR3. It uses the Baryonic model B2 from \cite{deSalas2019} (hereafter, named `Bary1') and an Einasto profile \citep{Einasto1965,Retana2012} for the DM halo. Model B is made by adding a circumgalactic medium (CGM) to model A. The CGM density profile has been implemented following \cite{Wang2019}, which is based on their modelling of the Magellanic Stream (see their Fig. 1). We verified that adding the CGM does not affect the rotation curve and its detected Keplerian decline because of the very low density of the ionised gas. Models C and D are taken from \cite{Jiao2021} and fit the rotation curve from \textit{Gaia} DR2 \citep{Eilers2019}. These two MW potential models use Model I from \cite{Pouliasis2017} (hereafter, named `Bary2') for the baryons and an Einasto potential \citep{Retana2012} for the DM halo. Finally, model E is the MW potential model adopted by \cite{Eilers2019} to reproduce the \textit{Gaia} DR2 rotation curve. The baryon model is Bary2 and the associated DM halo of the model E is based on a Navarro, Frenk, and White (NFW) profile \citep{Navarro1997}. The $M_{tot}$ values of these five potential models span a range of 2.06 to 8.14 $\times10^{11}\mathrm{M_{\odot}}$. Higher-mass models have shallower DM halo profile, resulting in higher $R_{200}$ values, so they can still fit the rotation curve from 5 to 30 kpc.

\begin{figure*}[t]
\centering
\includegraphics[width=0.8\linewidth]{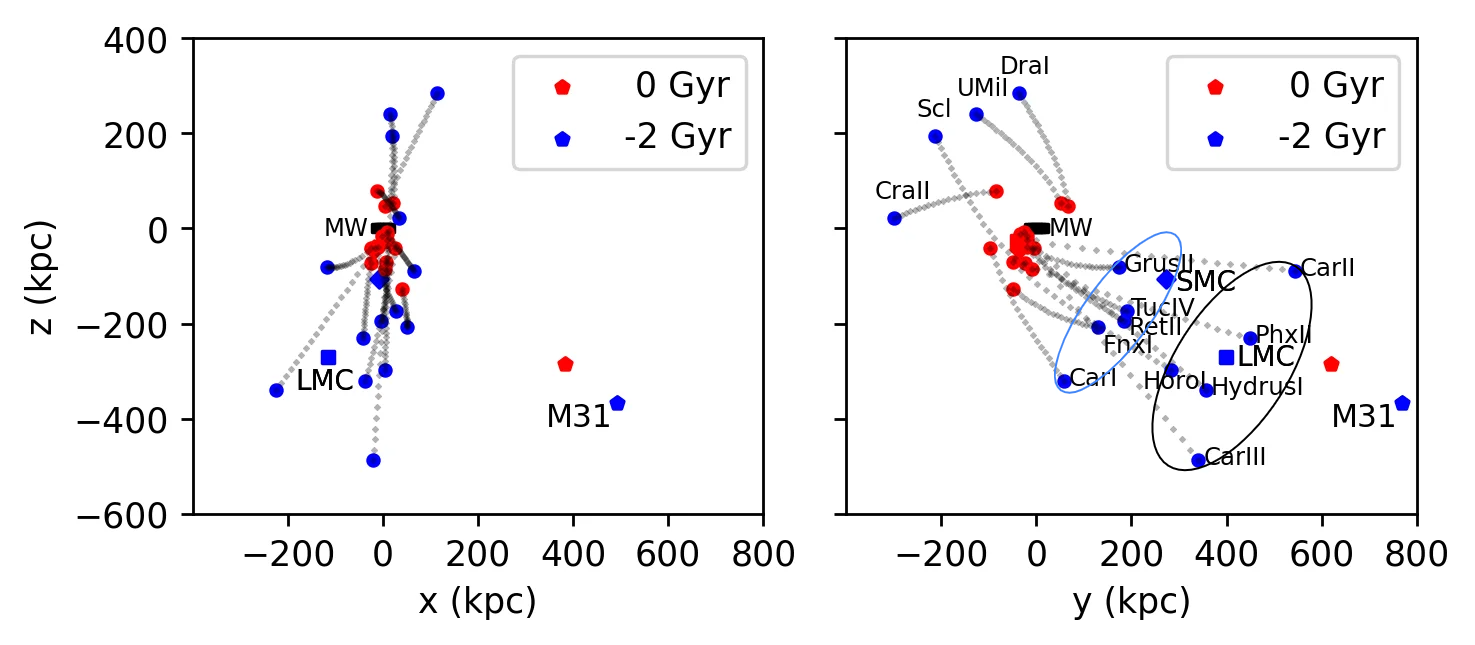}
 
\caption[Dwarf Traceback]{Traceback of the VPOS dwarf galaxies with mean proper motion from \cite{Li2021} and \cite{Luri2021} under MW potential model B in a left-handed Galactocentric coordinate frame. The MW is the black line at (0,0,0). Red and blue correspond to the positions at the current epoch and at -2 Gyr, respectively. Black and blue ovals indicate the two groups, LMC dwarfs and slow dwarfs, respectively.}
\label{DwarfTracebackB}
\end{figure*}

Our choice for the VPOS dwarfs comes from combining the analysis of \cite{Li2021} and \cite{Taibi2024}\footnote{The VPOS dwarfs considered in this paper: LMC, Small Magellanic Cloud, Carina, Carina II, Carina III, Crater II, Draco, Fornax, Grus II, Horologium I, Hydrus I, Phoenix II, Reticulum II,  Sculptor, Tucana IV, Ursa Minor.}. We only considered the dwarfs that are identified as VPOS dwarfs by both of these papers. The only exception to this criterion is Carina II, which \cite{Taibi2024} note as just outside the VPOS, but we kept it since it has been identified as an LMC satellite \citep{Erkal2020}. We used the galpy \citep{Bovy2015} package in Python for the orbit calculations. Within galpy, the objects are taken as point masses and are only affected by the MW potential. Any interactions with other dwarfs, MW gas, tidal effects, or dynamical friction were not included. Figure~\ref{DwarfTracebackB} shows the traceback with mean proper motion of dwarfs from \cite{Li2021} (dwarf galaxies) and \cite{Luri2021} (Magellanic clouds) under the MW potential B. 12 out of 16 VPOS dwarfs appear to be in their first infall to the MW halo following a similar general direction (see the y-z plane of Fig. \ref{DwarfTracebackB}) and are close to their pericentre as suggested by \cite{Taibi2024}. We notice that the current position of M31 is in a similar direction, although the dwarf orbits do not point exactly towards it in the x-axis (Fig. \ref{DwarfTracebackB}, x-z plane). For higher mass MW models, most dwarfs are bound because their orbital energy is not sufficient to escape a heavy potential, as shown in Appendix \ref{sec: Traceback MW E} for model E. In these MW models, only dwarfs with very high orbital energies can have trajectories that suggest a first infall. However, at a similar look-back time, they are found at a smaller distance from the MW, and their trajectories are more curved than for the low-mass MW models.    

According to the trajectories of Fig.~\ref{DwarfTracebackB}, the dwarfs can be classified into three groups. 
\begin{itemize}
  \item Six dwarfs with high orbital energy, including the LMC and its satellites \citep{Erkal2020}, hereafter called the `LMC dwarfs', whose velocities allow them to cover most of the distance between M31 and the MW within 2 Gyr (shown in black oval).
  \item A second group (six dwarfs including the Small Magellanic Cloud \footnote{SMC and its orbit has been significantly disrupted through an encounter with the LMC 0.2 - 0.3 Gyr ago \citep{Wang2019}. So, its traceback solely based on its current proper motion is not accurate.}) with a similar infall direction as the LMC dwarfs, hereafter called the `slow dwarfs' (blue oval), which may escape the MW but at a much lower velocity in our integrations.  
  \item A third group of four dwarfs appears to approach the MW from the opposite direction. These will be referred to as the `other dwarfs'. 
\end{itemize}

Our choice of dwarf traceback in galpy without any other gravitational interaction (except that of the MW) or gas interactions is not accurate. In the frame of a first infall scenario of VPOS dwarfs argued by \citet{Hammer2023}, most of their progenitors were gas-rich and DM-poor or free before their infall. Here, we suggest that the slow dwarfs of Fig. \ref{DwarfTracebackB} have had their velocity reduced by the ram pressure effects caused by the MW hot gas, and this facilitated their capture by the MW. This mechanism has been well illustrated by \citet[see the top-left panel of their Fig. 3]{Wang2024}, who showed a simulated Sculptor gas-rich progenitor arriving with a very eccentric orbit that was subsequently captured by the MW due to the ram pressure slowdown. On the other hand, the LMC is much more massive, implying that it is expected to be far less affected by the gas \citep{Wang2019}. This can be explained by the fact that the ram pressure is proportional to the surface, while the gravity is proportional to the mass and then to the volume. Hence, for a large body like the LMC, with a mass hundred to thousand times that of a dwarf spheroidal galaxy, the ram pressure slowdown is expected to be much less efficient. Thus, it should keep a very eccentric orbit \citep{Kallivayalil2013} even after reaching its pericentre. Furthermore, although the LMC dwarfs are affected by the ram pressure, the gravitational pull of the much more massive LMC would drag them along its orbit. So, for the LMC and its satellite dwarfs \citep{Erkal2020}, the ram pressure slowdown in their orbits is not as significant as the slow dwarfs.
\begin{figure}[t]
\centering
 \includegraphics[width=0.9\columnwidth]{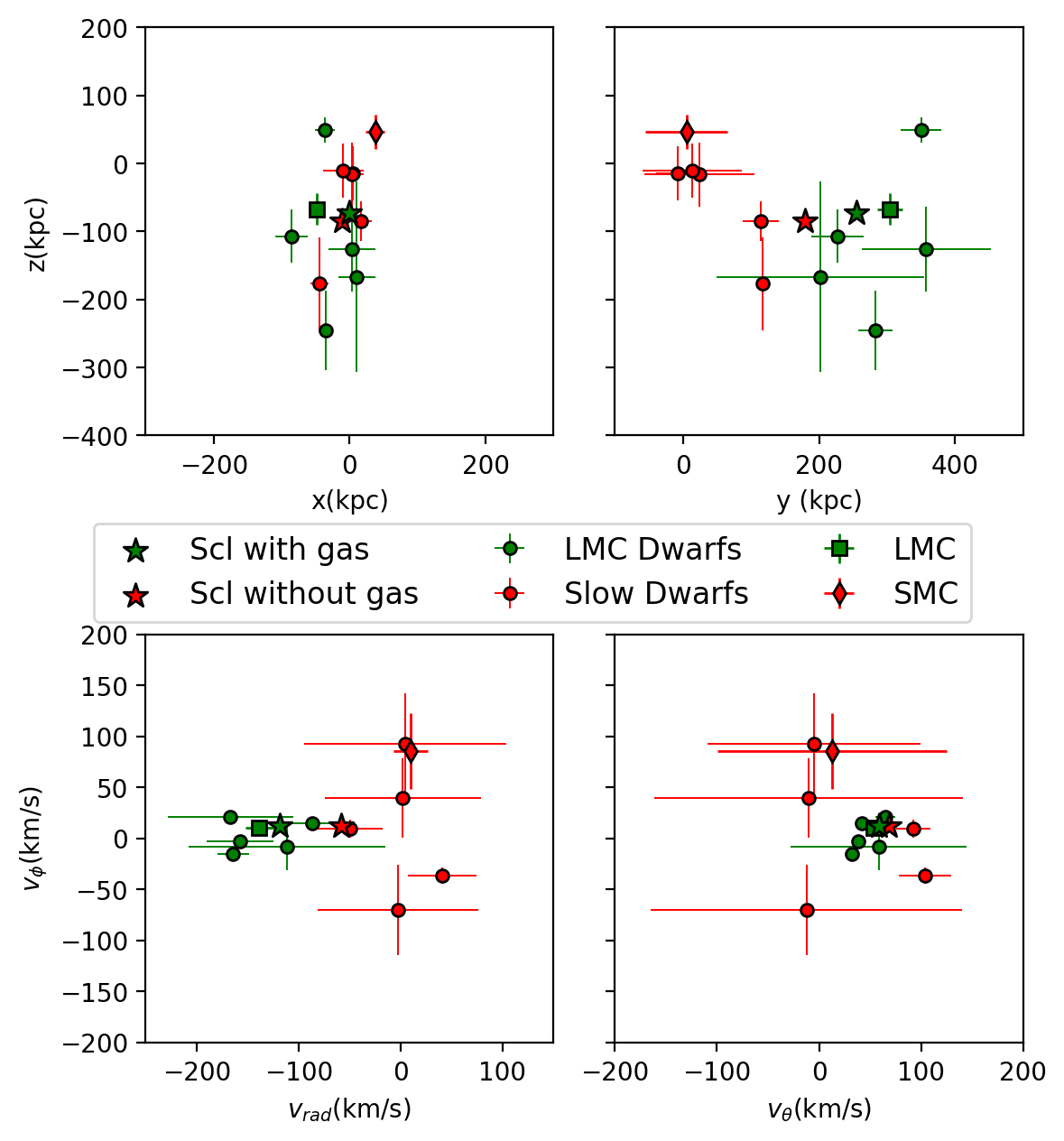}
 \caption[LMC dwarfs vs slow dwarfs]{ LMC (green square), LMC dwarfs (green circle), slow dwarfs (red circle), and Sculptor-like dwarf infall with gas (green star) by \cite{Wang2024} and its traceback without correcting for gas (red star) with galpy in 6D space at -2 Gyr.}
 \label{LMC dwarfs vs slow dwarfs}
\end{figure}

To verify this hypothesis, we took the current position and velocity from the simulation of the infall of a Sculptor-like dwarf by \cite{Wang2024} and traced it back in galpy using the same MW potential model, with the assumption of point mass and that it is only affected by the MW gravity, (MW mass = $6.9\times 10^{11} M_{\odot}$) used by \cite{Wang2024}. We also traced back the LMC dwarfs and the slow dwarfs in this potential and roughly aligned the Sculptor orbit with these dwarfs. Figure \ref{LMC dwarfs vs slow dwarfs} shows the comparisons in 6D when the dwarfs are traced back in time by 2 Gyr. LMC dwarfs, slow dwarfs, Sculptor with gas and Sculptor without gas occupy the same spatial position in the $x, z, v_{\phi}, v_{theta}$ coordinates. The main difference is in radial velocity, $v_{rad}$, and subsequently in the y-axis since the radial velocity direction of these dwarfs is close to the y-axis (Fig. \ref{DwarfTracebackB}). The hydrodynamical simulation of Sculptor gas-rich progenitor shows that its initial radial velocity is consistent with that of the LMC dwarfs. On the other hand, its traceback in galpy as a point mass and without gas is consistent with the slow dwarfs. The difference in the y-axis for the two Sculptor simulations is also comparable to the difference between the two dwarf groups. This suggests that if we could account for the ram pressure slowdown, the slow dwarfs would occupy a similar location in 6D space as the LMC dwarfs during the infall. Hence, for the present study, the LMC and its satellite dwarfs are the best-suited targets for an analysis of their origin within the scope of galpy, and in the following section we consider the LMC as the representative of all the 12 dwarfs within the groups LMC dwarfs and slow dwarfs.

\section{6D association}
\label{sec:6D association}

\begin{figure}[h!]
\centering
 \includegraphics[width=\linewidth]{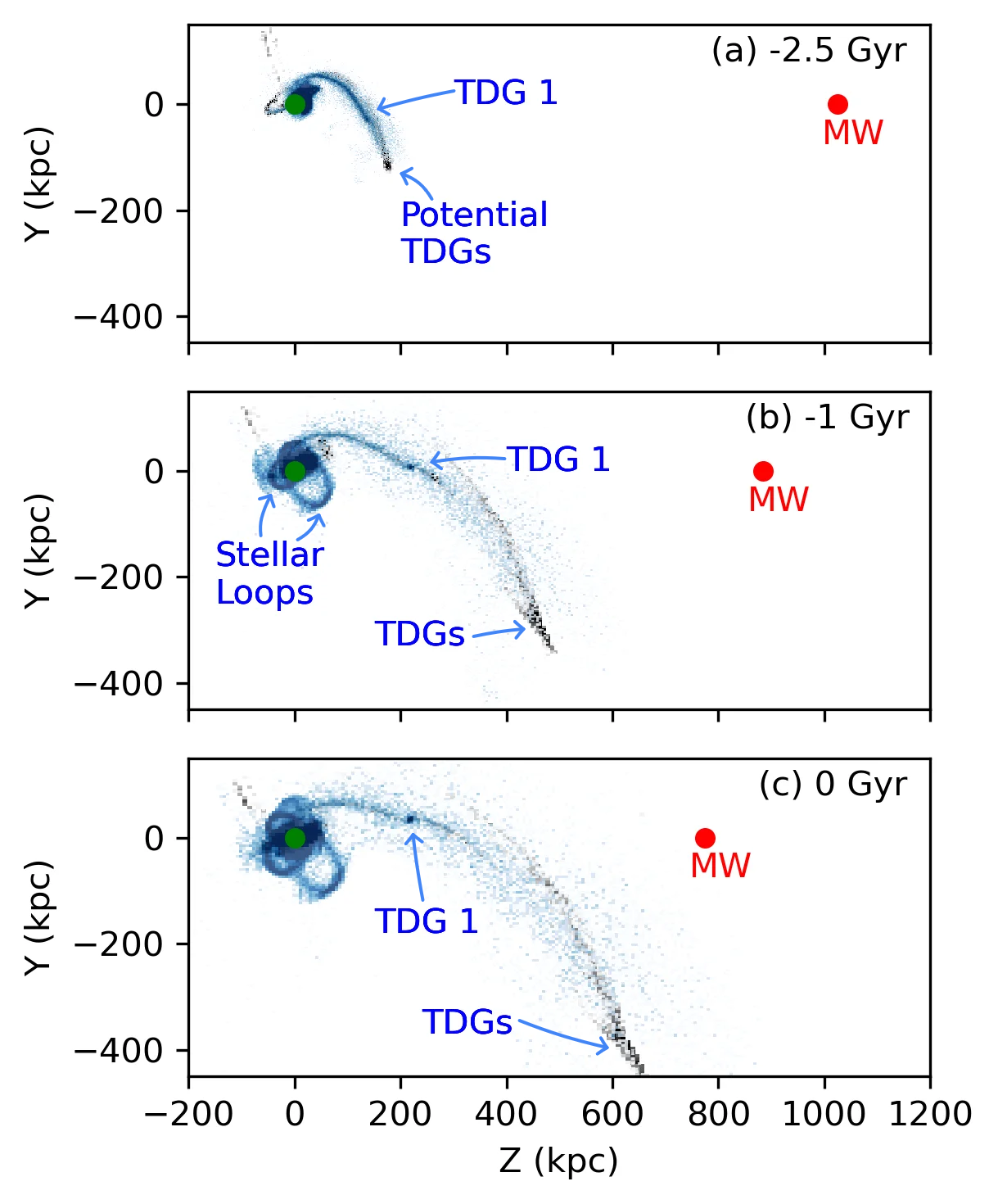}
 \caption[M31 tidal to MW]{Movement of tidal tail (star particles in blue and gas particles in grey) from M31 merger model 288 towards the MW and formation of TDG within the tail. Panel a: Tidal tail formed during the second passage of M31 merger with some zones of over-density. Panel b: Particles of the tidal tail that are close to the M31 fall back and form stellar loops that ultimately form the GSS. Panels b and c: Distant parts of the tail escape M31, which then expand moving towards the MW. The formation of at least one TDG and potentially a few more in the dense zone can also be identified. In this coordinate system, the M31 centre (green) is fixed at (0,0,0), and the MW is at 0 Gyr is on the z-axis. The stars and gas particles are shown in log-normal density to identify the over-dense regions. The position of the MW (red dot) is calculated for a radial approach between the two galaxies taken as the point mass for the MW mass of model B and the M31 mass from \cite{Hammer2025}. The MW's gravitational impact on the particles is not considered.}
 \label{M31 TT2 to MW}
\end{figure}

We studied the tidal tail particles generated by each of the five M31 merger models presented in \cite{Hammer2018}. All these models predict a tidal tail (originating from the second passage of the merger, 2-3 Gyr ago) passing near the MW, and the goal of this paper is to compare its 3D position and 3D velocity with those of the LMC and related dwarfs. Figure \ref{M31 TT2 to MW} illustrates the evolution of this tidal tail and the formation of TDGs. Due to the abundance of stellar material and gas during the final merger stage, some TDGs can form in a high-density tidal tail \citep{Wetzstein2007}, and a few of them can be seen in the figure. Part of the tail that is close to M31 is bound by its gravity and falls back to form stellar loops that we can observe as the GSS\footnote{We note that both the GSS and tidal tail content are essentially made of material from the secondary progenitor galaxy. Both share the same angular momentum direction, which is related to the merger orbital plane. Since the GSS is seen edge-on from the MW, it implies that the MW is within the orbital plane of the M31 merger.}. With time, the other part of the tail at large distances from M31 is not bound to the merger remnant, while the TDGs continue to gather more and more material. The second passage occurred 3 Gyr ago, and later the tail is travelling towards the MW.

\begin{figure*}[ht]
\centering
 \includegraphics[width=0.8\linewidth]{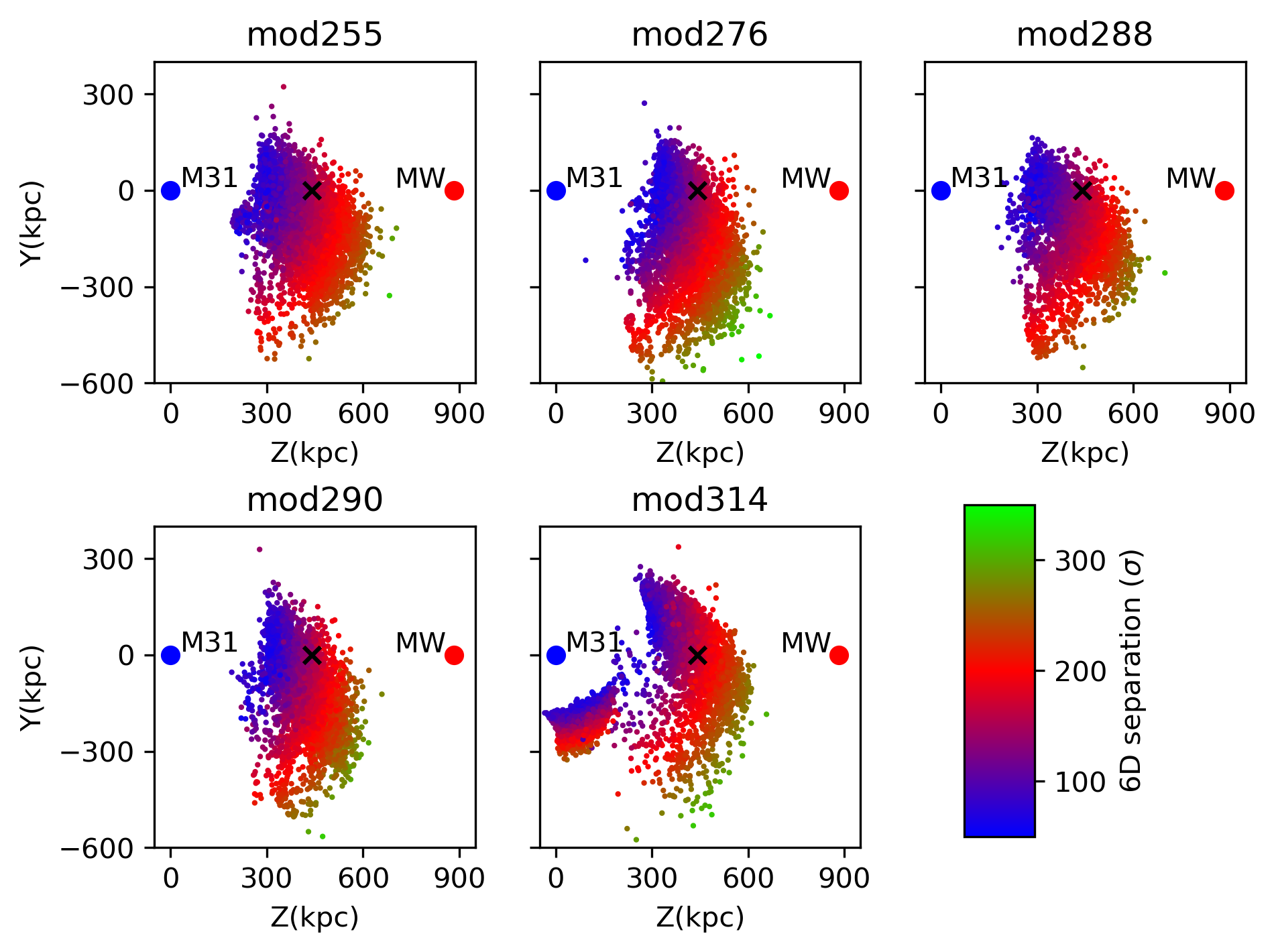}
 \caption[M31 Merger Models]{Selected stellar particles in the tidal tails and their 3D velocities (in colour) from the M31 merger models at -1 Gyr. The M31 is fixed at (0,0,0), and the MW at 0 Gyr is on the z-axis. The position of the MW (in red dot) is calculated for a radial approach between the two galaxies taken as the point mass for the MW mass of model B and the M31 mass from \cite{Hammer2025}. The x-marks denote the midpoint.}
 \label{merger_mod}
\end{figure*}

However, M31 merger models do not account for the MW potential. We show in Appendix \ref{sec: MW on M31 merger} that the effect of the MW is small during the 2nd passage of the merger that generates this tidal tail. Furthermore, we only took the stellar particles of the tidal tails that are at a distance of more than 250 kpc and are moving away from the M31 centre. These particles are unbound to M31, whereas, the particles close to M31 are bound and fall into M31 to create the GSS \citep{Hammer2018}. We chose the time such that their median location is about halfway between M31 and MW. This occurred at a look-back time corresponding to about 1 Gyr. Using galpy, these particles would only be affected by the potential of MW, while their initial velocities are determined by the M31 merger model. We tested the effect of the M31 potential in the MW neighbourhood by implementing it as a moving object potential in galpy along with the MW potential and found that the LMC orbit has a change of less than 2\% within this distance scale (see Appendix \ref{sec: M31 Potential}). Hence, the M31 potential could be ignored in the following calculations. The positions of the stellar particles of the tidal tails used in our calculation are shown in Fig. \ref{merger_mod}.

Instead of the current epoch, we compared the tail particles with the dwarf positions at a look-back time of 1 Gyr. This choice was made because if we let the tail particles come too close to the MW, we would have to account for the effect of the ram pressure, which is beyond the scope of this paper, and because the galpy package cannot account for hydrodynamical physics (i.e. ram pressure effects). However, as demonstrated in Fig. \ref{LMC dwarfs vs slow dwarfs}, by comparing with the Sculptor infall with and without gas, the massive LMC and its satellites are much less affected by the MW gas. So, using galpy, at -1 Gyr and $\sim$250 kpc away from the MW centre, we will have good estimates for both the LMC dwarfs and the tail particles. On the other hand, if we calculate at current time (0 Gyr), we will still find the same result (since the galpy orbits are only under the MW gravity), but the estimates for the position and velocity of the tail particles would be much more uncertain. Furthermore, these dwarfs are close to their pericentres at the current time. So, the change in position and velocity is rapid, and it would require a much higher time resolution for comparisons at the current epoch. For the LMC and its satellite dwarfs, we corrected their position and velocity by accounting for the recent LMC--Small Magellanic Cloud (SMC) encounter, dynamical friction, and ram pressure using the analysis of \cite{Hammer2015}. As shown in Appendix \ref{sec:LMC-SMC collison}, the change in the LMC orbit is added as a small shift at every time step. However, since the SMC is much lighter, it has been significantly disrupted by this encounter \citep{Wang2019}.

The timing of the M31 merger models has an uncertainty of about 1 Gyr \citep{Hammer2018} due to the uncertainty of the precise time during which the simulated post-merger galaxies have properties reproducing the M31 observations. We let the time of tidal tail infall be a free parameter and vary from 0 to 2.5 Gyr with gaps of 0.05 Gyr. Another important parameter is the proper motion of M31, which is very poorly constrained. For example, \cite{vanderMarel2019}, \cite{Salomon2021}, and \cite{Rusterucci2024} report M31 proper motion from \textit{Gaia} DR2, EDR3, and DR3, respectively, with large uncertainties, especially when the \textit{Gaia} systematics are properly included. \textit{Gaia} EDR3 systematics of 25 \textmu as yr$^{-1}$ would correspond to a 1$\sigma$ uncertainty of 90 km s$^{-1}$ at the M31 distance \citep{Lindegren2021}. So, in our analysis, we considered the M31 proper motion to be a free parameter. We used a uniform grid (with gaps of 1 \textmu as) of the M31 proper motion values $(\mu_{\alpha}, \mu_{\delta})$ = (-20 to 110, -105 to 25) \textmu as yr$^{-1}$, which includes the previous estimates.

Our goal is to compare the 6D positions and velocities ($ x_i = [x,y,z, v_{rad}, v_{\theta}, v_{\phi}]$ in the left-handed Galactocentric coordinates) of the tidal tail particles with those of the observed back-traced dwarfs. For a given set of MW potential model, M31 merger model, proper motion, and tidal tail infall time, we calculated the position and velocity of these dwarfs and the tidal tail particles. The size of the tidal tail is very large ($\sim$ 1 Mpc), compared to the distance between the LMC dwarfs at -1 Gyr ($\sim$ 100 kpc). Therefore, we selected those tidal tail particles that are close to the LMC and the LMC dwarfs in the 6D space. We set the criteria based on the means and standard deviations of the dwarf position and velocity coordinates. We followed several steps, as described below:
\begin{enumerate}

\item For each of the six dimensions (i = 1 to i = 6), we calculated the weighted averages, $\bar{x}_{i,\ \mathrm{dwarf}}$, and the weighted standard deviations, $\sigma_{i,\ \mathrm{dwarf}}$, of the LMC dwarfs.

\item We selected the 6D space within five standard deviations of the weighted average ($\bar{x}_{i,\ \mathrm{dwarf}} \pm 5\sigma_{i,\ \mathrm{dwarf}}$). We also added a further constraint of two standard deviations of the 3D separation and velocity difference ($\bar{r}_{\mathrm{dwarf}} + 2 \sigma_{r, \ \mathrm{dwarf}}$ and $\bar{v}_{\mathrm{dwarf}} + 2 \sigma_{v, \ \mathrm{dwarf}}$). The tidal tail particles that occupy this 6D space are called the `dwarf neighbours.' We put a tighter constraint on the 3D separation and 3D velocity difference to constrain the kinetic and potential energy of the tidal tail particles to match that of the LMC dwarfs.

\item On the dwarf neighbours, we put a further 2$\sigma$ constraint on the 3D separation and velocity difference based on the LMC ($\bar{r}_{\mathrm{LMC}} \pm 2 \sigma_{r,\ \mathrm{LMC}}$ and $\bar{v}_{\mathrm{LMC}} \pm \sigma_{v,\ \mathrm{LMC}}$) to enforce similarity with the LMC's orbital energy. These tidal tail particles are called the `LMC neighbours.' 

\item If there are ten or more LMC neighbour particles, we consider that as a matching solution. We calculated the `6D separation' between the LMC and LMC neighbours to compare how far in 6D space they are from each other. The 6D separation is defined as the root mean square deviation (RMSD) calculated over six dimensions with the quadratic sum of uncertainties from the LMC and the LMC neighbours.\\

$\mathrm{6D}\ \mathrm{separation} = \sqrt{ \dfrac{1}{6} \smashoperator[r]{\sum_{n=1}^{6}}\left( \dfrac{\bar{x}_{i, \ \mathrm{ LMC}} - \bar{x}_{i, \ \mathrm{LMC}\ \mathrm{Neighbours}}}{ \sqrt{\sigma^2_{i, \ \mathrm{LMC}} + \sigma^2_{i, \ \mathrm{LMC}\ \mathrm{Neighbours}}}} \right)^2 }$
\end{enumerate}

\begin{table*}[!tb]
\centering
\begin{threeparttable}
\caption{\label{Table result}Parameters for the best solutions of each of the MW potential models and M31 merger models.}
\begin{tabular}{cccccc}
\hline
\hline
 MW potential & M31 merger model & M31 pm (\textmu as/yr)  & Time difference (Gyr) & 6D separation ($\sigma$) & LMC neighbours\\
 \hline
  \multirow{5}{*}{Model A}&mod255 &83, 6 &1.25 &0.49 &27\\
  &mod276 &64, 9 &1.05 &0.80 &68\\
  &mod288 &80, 3 &1.15 &0.65 &55\\
  &mod290 &84, 4 &1.20 &0.35 &48\\
  &mod314 &67, -12&0.9 &1.00 &12\\
 \hline
  \multirow{5}{*}{Model B}&mod255 &87, -7 &1.4 &0.46 &17\\
  &mod276 &71, 5 &1.25 &0.78 &44\\
  &mod288 &70, 2 &1.2  &0.64 &52\\
  &mod290 &83, -1&1.25 &0.38 &54\\
  &mod314 &60, -9&1.05 &0.98 &13\\
 \hline
  \multirow{5}{*}{Model C}&mod255 &70,-22 &1.9 &0.70 &18\\
  &mod276 &57, -13 &1.9 &0.75 &26\\
  &mod288 &56, -17 &2.1 &0.79 &19\\
  &mod290 &73, -20 &1.9 &0.57 &20\\
  &mod314 &43, -35 &1.9 &1.22 &21\\
 \hline
  Model D&mod276 &40, -32 &2.1 &4.03 &12\\
 \hline
\end{tabular}
\begin{tablenotes}
\item Only the M31 merger model 276 has solutions under MW potential model D, and none of the merger models have a solution within this calculation under the MW potential model E.
\end{tablenotes}
\end{threeparttable}
\end{table*}

We chose the 6D separation as the main metric to quantify how consistent the tidal tail particles are with the LMC and related dwarfs. This assumes that the individual particles are a realistic representation of the overall phase space occupied by the tidal tail. The number of LMC neighbour particles is considered as a secondary metric. The best solution for a certain MW potential model and M31 merger model is characterised by having the highest number of the LMC neighbours within 0.05$\sigma$ of the lowest 6D separation value. Table \ref{Table result} presents parameters for the best solutions. Low-mass MW models (A and B) provide the best solutions: three out of the five M31 merger models have a solution with a 6D separation of less than 0.65 $\sigma$. MW potential model B has a slightly higher mass than model A due to the addition of the CGM, but the 6D separation values are similar. The time difference between the dwarfs and tidal tail particles is 0.9 to 1.4 Gyr depending on the chosen M31 merger model. Considering the 1 Gyr uncertainty expected for the M31 merger models, this time gap is consistent with a simultaneous occurrence. Some parts of the time gap can arise from the slowdown due to the ram pressure for both the LMC and the tidal tail, which could not be accounted for by our calculation. The intermediate-mass MW model C has a slightly higher 6D separation compared to A and B. However, the time gap required for such solutions is around 2 Gyr, which is too large to preserve the reproduction of the M31 properties by a post-merger model \citep{Hammer2018}. For the high-mass MW model D, we find only one solution with the M31 merger model 276, which, however, provides too large values for both the time gap and 6D separation. Finally, the high-mass MW model E \citep{Eilers2019} does not provide any satisfying matching solution for any M31 merger models. Within the tidal tails from the 5 M31 merger models of \citet{Hammer2018}, model 290 usually gave the lowest values of the 6D separation and the largest number of LMC neighbour particles, which leads to the best agreement. On the other side is the M31 merger model 314 with a high value of 6D separation and the lowest number of LMC neighbour particles.

\begin{figure}[ht!]
\centering
 \includegraphics[width=\columnwidth]{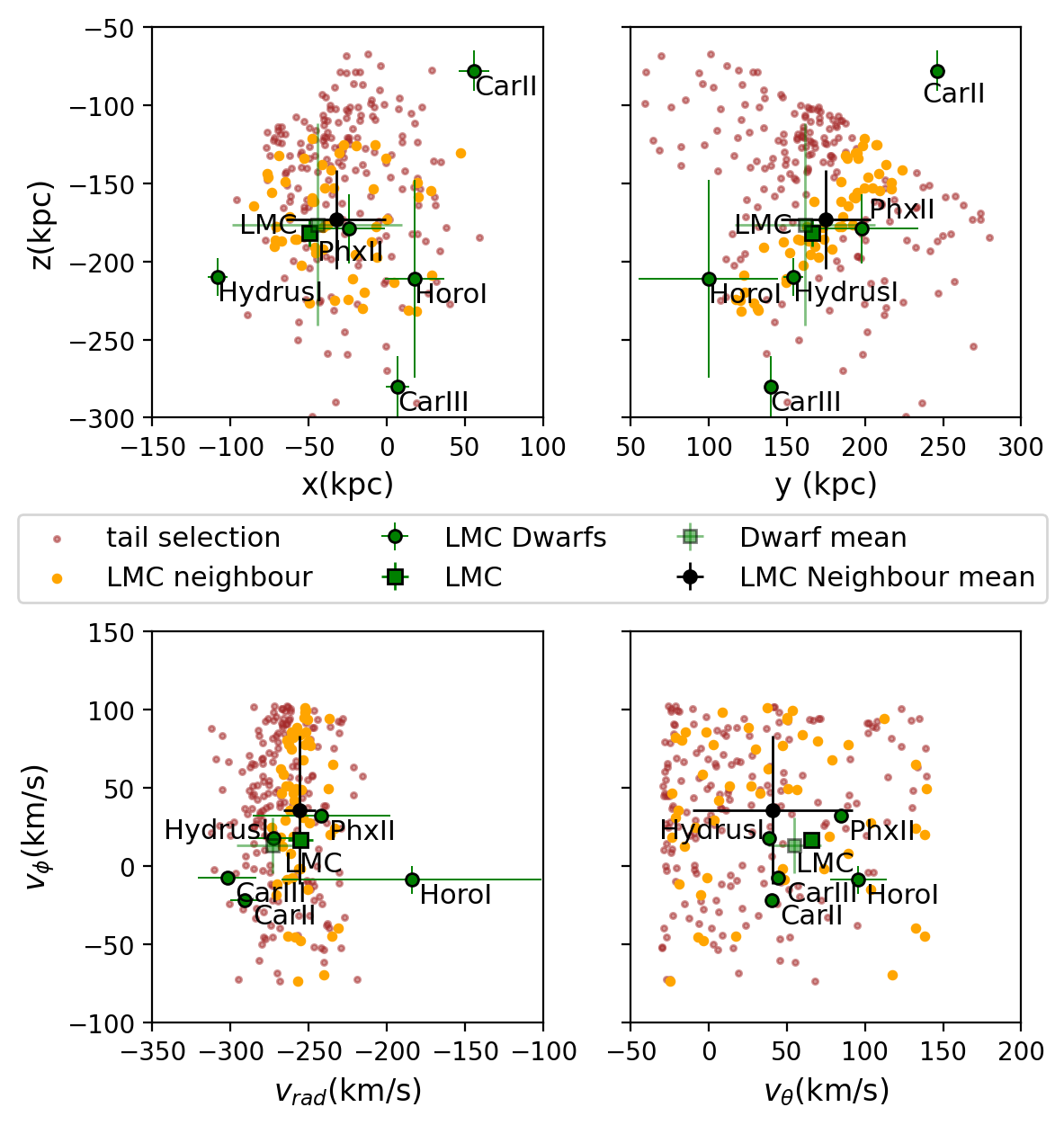}
 \caption[model 290, 6D association]{Best solution for M31 merger model 290 under MW potential model B at -1 Gyr. The 6D separation between the LMC and its neighbouring particles is 0.38$\sigma$.}
 \label{modelB model290}
\end{figure}

Figure \ref{modelB model290} shows one of the best solutions of this analysis,  which occurred for the MW mass model B and the M31 merger model 290. The LMC and the weighted mean of the dwarfs are very close in 6D phase space, which indicates that the LMC can indeed be used as a proxy for all the LMC-related dwarfs. Across all six dimensions, the LMC and the LMC neighbours are separated by less than 0.5 sigma, and the LMC is well surrounded by the LMC neighbour particles. Except for the spatial location of Carina II (see the top-right panel of Fig. \ref{modelB model290}), the LMC dwarfs also have neighbouring particles in all 6D within their error bars. So, the LMC and related dwarfs could be associated with parts of this tidal tail in 6D. 

\begin{figure}[!ht]
\centering
 \includegraphics[width=\columnwidth]{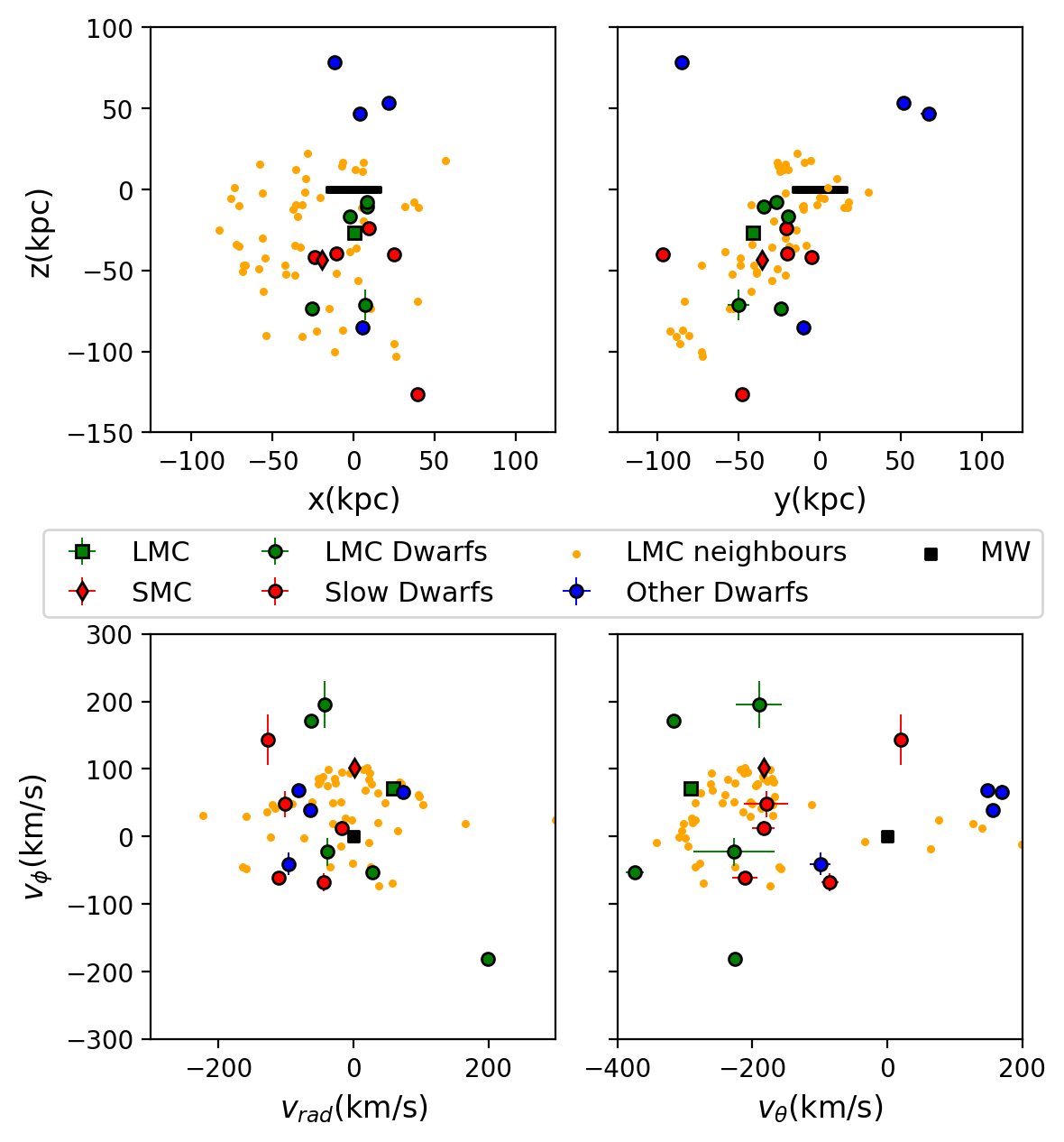}
 \caption[VPOS reproduced]{Best solution of M31 merger model 290 under MW potential model B. LMC neighbour particles (yellow) are at 0.1 Gyr for the best association. The LMC dwarfs (green), slow dwarfs (red), and other dwarfs (blue) are at the current epoch (0 Gyr). The MW is shown in black.}
 \label{vpos reproduced}
\end{figure}

The association between parts of the tidal tail and LMC-related dwarfs has been identified when the dwarfs are at -1 Gyr. If we let these LMC neighbour particles fall into the MW, they indeed arrive at a similar place in 6D as the current observation of LMC dwarfs, since within galpy these particles and dwarfs are only affected by the MW gravity. For example, we took the solution shown in Fig. \ref{modelB model290} and let the LMC neighbour particles evolve so that they reach near the current LMC at 0.1 Gyr. We allowed this (small) time gap for the best association and it could be associated with ram pressure slowdown. Figure \ref{vpos reproduced} shows the dwarfs and tidal tail particles in the 6D space. We also included the other dwarfs. The LMC neighbours are mostly consistent with the LMC, especially in position space. The top two panels (position space) show the alignment of VPOS dwarfs in a line almost perpendicular to the MW disc. Most of the other dwarfs are situated on the upper side of the MW disc (the positive side of the z-axis). The contrast is also seen in $v_{\theta}$ as most of the other dwarfs have positive values at 0 Gyr. In the other four dimensions, there are no significant differences among these dwarf groups.

\begin{figure}[!h]
\centering
 \includegraphics[width=\linewidth]{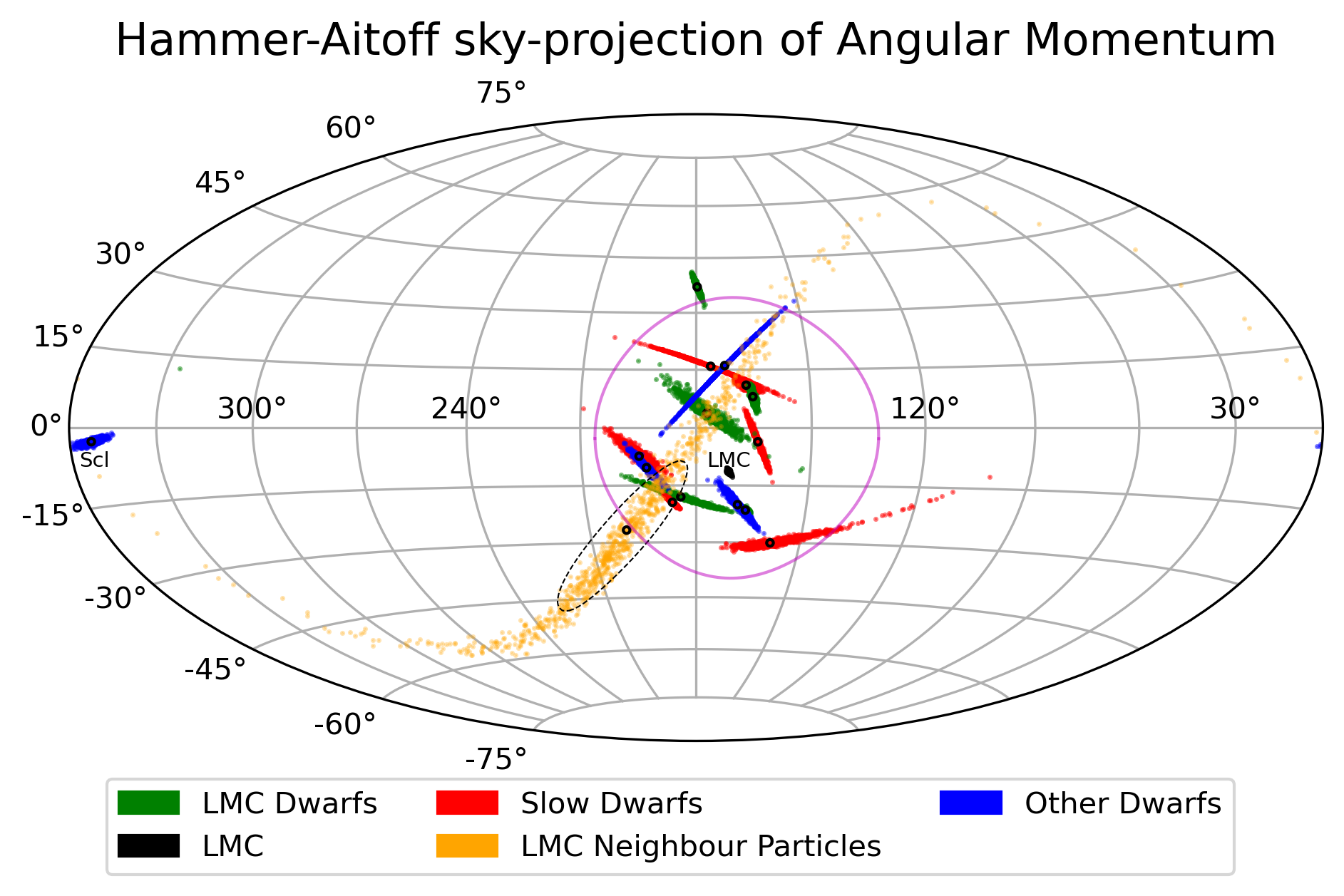}
 \caption[Angular Momentum Pole]{Angular momentum poles of the LMC neighbour particles (the best solution of M31 merger model 290 and MW potential model B) and the VPOS dwarfs with 2000 Monte Carlo simulations. The small black circles show the median, the dashed black oval the 1$\sigma$ spread, and the magenta circle the VPOS location by \cite{Fritz2018}. Sculptor has a counter-rotating orbit.}
 \label{Angmom_pole}
\end{figure}

Another way to identify the VPOS dwarfs is by using their angular momentum direction, which is almost perpendicular to the rotation of the MW \citep{Metz2008, Pawlowski2011, Fritz2018}. We made a similar sky projection (Fig. \ref{Angmom_pole}) by performing Monte Carlo simulations for both dwarfs and LMC neighbour particles. It shows that the angular momentum poles for the LMC neighbour particles are broadly consistent with VPOS dwarfs. The spread of the angular momentum pole for the LMC neighbour particles is much higher than the VPOS dwarfs because the error bar of their mean in all of 6D is much bigger. In the calculation of the 6D separation, the errors in the six dimensions are averaged (RMSD), while in the calculation of the angular momentum direction, the errors are accumulated. So, although the LMC is located within the LMC neighbour particles and the 6D separation is only 0.38$\sigma$ (Fig. \ref{modelB model290}), the angular momentum poles are separated by more than 1$\sigma$ (Fig. \ref{Angmom_pole}).

\begin{figure*}[!ht]
\centering
    \includegraphics[width=.8\linewidth]{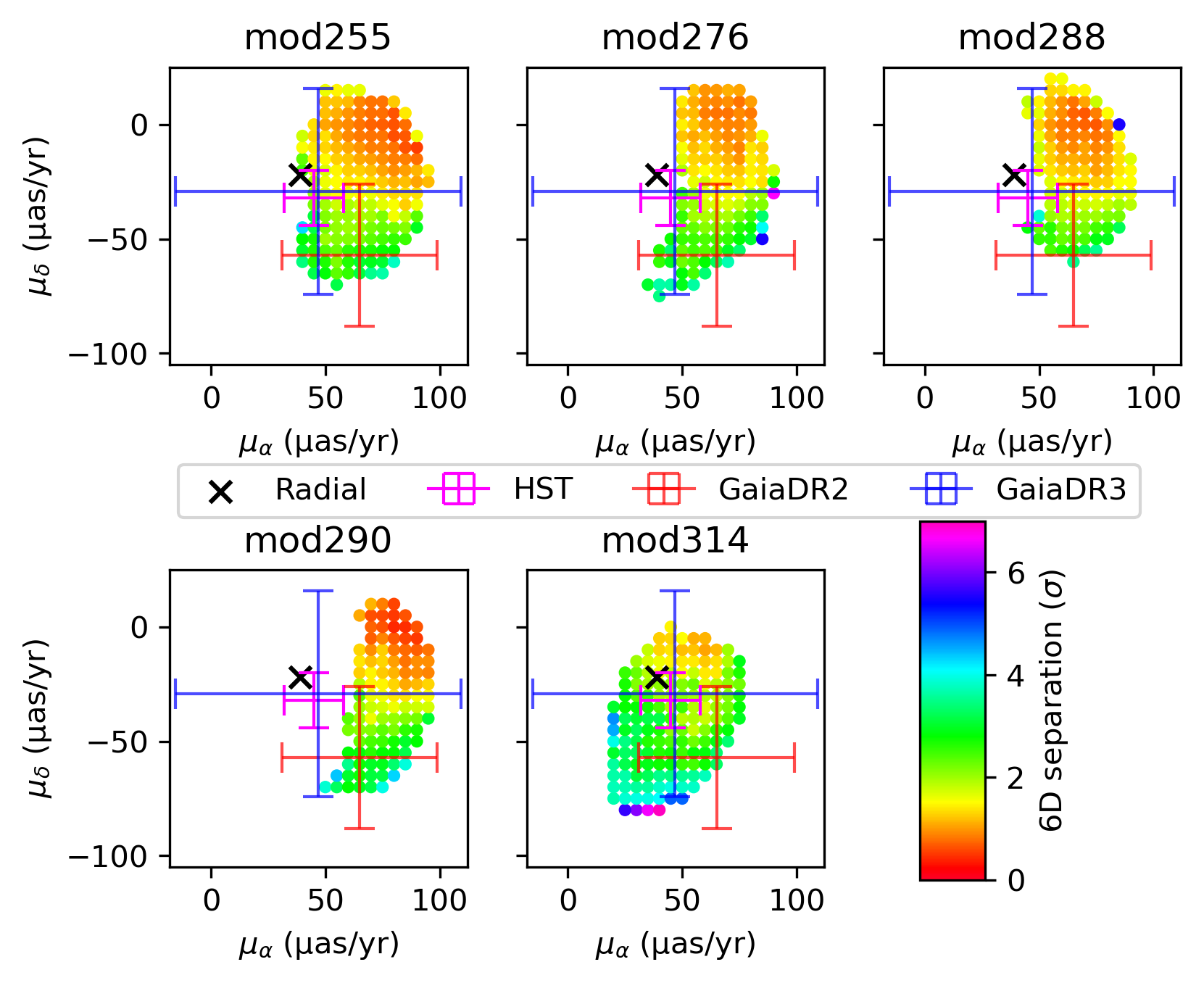}

 \caption[PM grid]{6D separation values for different M31 proper motion values with the M31 merger models under MW potential model B. Proper motion values without colours result in fewer than ten LMC neighbour particles. The M31 proper motions for a radial approach to the MW and derived from HST, \textit{Gaia} DR2, and \textit{Gaia} DR3 data are also shown.}
 \label{pm Grid}
\end{figure*}

Figure \ref{pm Grid} shows how the 6D separation value depends on the M31 proper motion under the MW potential B. The 6D separation is highly dependent on the M31 proper motion because it affects both the direction and the speed of the tidal tail to the MW. Some of the proper motion values yield no solution (if there are fewer than ten particles in the LMC neighbourhood), while the best solutions are found around $(\mu_{\alpha}, \mu_{\delta})$ = (75$\pm$10, $\pm$10) \textmu as. The values that result in low 6D separation are consistent with the M31 proper motion found by \cite{Rusterucci2024}, who included a thorough analysis of the \textit{Gaia} systematics. Furthermore, it is also consistent within 1$\sigma$ of the M31 proper motion in the right ascension ($\mu_{\alpha}$) and 2$\sigma$ in the declination ($\mu_{\delta}$) found using the \textit{Gaia} DR2 \citep{vanderMarel2019}. Given the large uncertainties and systematics of the M31 proper motion, these values with low 6D separation (in red in Fig. \ref{pm Grid}) can be considered to be consistent with the observations.

\label{sec: discussion}
\section{Discussion}
We show that 12 out of the 16 VPOS dwarfs could be associated with the tidal tail particles originating from a recent major merger at the M31 location in terms of their 3D position and 3D velocity, assuming the low-mass MW models predicted by \textit{Gaia} DR3.  However, could a massive dwarf such as the LMC be a TDG? \cite{Kaviraj2012} performed an extensive search for the TDGs in the SDSS data through the Galaxy Zoo project. They compared the stellar mass ratio between the parent galaxy and the TDG (see their Fig. 6), which would be approximately 55 when comparing the LMC to M31 stellar masses. If associated with an ancient gas-rich merger that formed M31, and according to \cite{Kaviraj2012}, the LMC stellar mass would be consistent with the TDGs having such a mass ratio. Furthermore, \cite{Haslbauer2019} found TDG candidates in the Illustris-1 simulation with their most massive TDG having mass comparable to the LMC. The LMC mass is also consistent with the upper limits of the TDG mass found from the simulations in \cite{Bournaud2006}.

However, would classifying the LMC as a TDG be consistent with its substantial mass? The LMC mass estimates can be classified into three categories. In the first one, the LMC mass is derived from the internal kinematics of stars (M($<$ 8.7 kpc) = 1.7 $\pm$ 0.7 $\times$ $10^{10}$ M$_{\sun}$, \citealt{vanderMarel2014}) and of globular clusters (M($<$ 13.2 kpc) = 2.66 $\pm$ 0.4 $\times$ $10^{10}$ M$_{\sun}$, \citealt{Watkins2024}). In the second category, the LMC mass is deduced from its impact on the MW, and the most accurate result is coming from the disc sloshing \citep{Conroy2021,Erkal2021} caused by the first approach of the Magellanic system. \citet{Erkal2021} assumed the total mass of the MW to be 8.73 $\times$ $10^{11}$ M$_{\sun}$ and an LMC total mass of 1.5 $\times$ $10^{11}$ M$_{\sun}$ (i.e. a mass ratio of 17\%). \citet{Watkins2024} extrapolated their value using a shallow NFW profile and derived a similar total mass for the  LMC compared to that from \citet{Erkal2021}. In the third category, the LMC mass is estimated from the modelling of the Magellanic Stream, assumed to result from ram pressure caused by the Galactic coronae on the LMC and SMC neutral gas \citep{Hammer2015,Wang2019}, which requires a specific mass for the LMC in order to reproduce especially the HI-stream. The modelling reproduced both the HI and HII streams as well as the bridge. It leads to a maximal mass for the LMC of about 2 $\times$ $10^{10}$ M$_{\sun}$ \citep{Wang2022}. Interestingly, all these values can be in agreement if the MW mass is much smaller (see MW potential models A and B in this paper), and thus the nature (tidal dwarf vs DM-dominated) of the LMC is still open.

In Fig. \ref{pm Grid} and the relevant discussion, we focus on the \textit{Gaia} DR2 \citep{vanderMarel2019} and \textit{Gaia} DR3 \citep{Rusterucci2024} measurement of the M31 proper motion. We do not show \textit{Gaia} EDR3 proper motions \citep{Salomon2021} in the figure since it did not consider the systematic issues arising from the mismatch between measurements using red versus blue stars — these issues were addressed in \cite{Rusterucci2024}. Another available M31 proper motion comes from the \textit{Hubble} Space Telescope (HST) measurements \citep{vanderMarel2012}, which have a much smaller reported uncertainty compared to these \textit{Gaia} measurements. But this measurement used 3 small fields (compared to the size of the M31 in the sky), only one of which is on the M31 disc \cite[see their Fig. 1]{Brown2006}. Furthermore, they used an old model of the M31 with minor mergers to correct these three fields. These two issues would introduce significant uncertainty (and possible inconsistency with our past merger model of M31) that has not been accounted for. Hence, we stuck to the M31 proper motions only using the \textit{Gaia} data that sample a significant amount of stars in the M31 disc. \cite{vanderMarel2019} reported \textit{Gaia} DR2 + HST data in their work; however, the uncertainties in the combined value are dominated by the reported low uncertainty of the HST proper motion from \cite{vanderMarel2012}. Hence, we used their \textit{Gaia} DR2-only value in our work.

The LMC and the related dwarfs could be associated with the incoming tidal tail particles in the case of the low-mass MW models with baryon fractions higher than what is found in the literature for similar galaxies. But these low-mass models are supported by the \textit{Gaia} DR3 rotation curves \citep{Jiao2023, Ou2023} as well as the MW accretion history \citep{Hammer2024}. Establishing that the MW has a high mass would be one way to falsify our scenario. The M31 merger models have 20\% of baryons, which is similar to the cosmological fraction of 15.5\% found by Planck \citep{Planck2020}, but it also reproduces the observational features of the M31. In fact, to match the rotation curve of the M31, the M31 merger models need an even lower mass of $\sim 4.5 \times 10^{11} M_{\odot}$ within $R_{200}$ = 137 kpc \citep{Hammer2025}. The observational estimates of the total mass of the local group, such as $2.3\times10^{12} M_{\odot}$ from the local Hubble flow \citep{Penarrubia2014}, include all the mass within a 1-3 Mpc radius, whereas the estimates of the masses of the MW or M31 are for within $R_{200} \sim$150 kpc for these galaxies. The rest of the local group mass, within about 1000 times the volume of these galaxies, could be associated with the DM, CGM, and intergalactic medium outside the $R_{200}$ of the galaxies.

\begin{figure*}[!htp]
\centering
    \includegraphics[width=\linewidth]{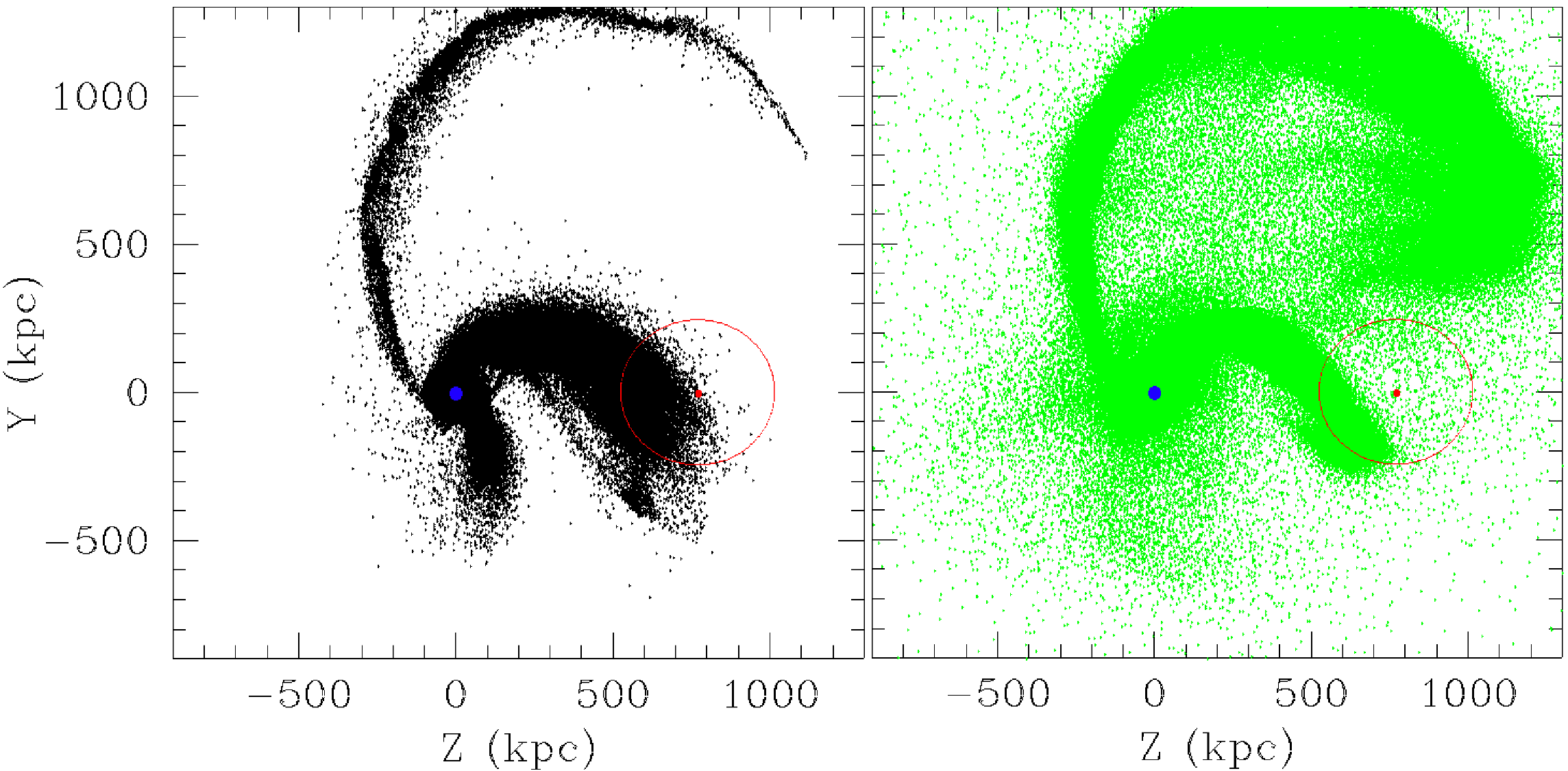}

 \caption[Other Tails]{Stars (black dots in the left panel) and gas (green dots in the right panel) from the M31 merger model 336 with 24 million particles, that shows the formation of more than two tidal tails. The M31 is shown with a blue dot and the MW (red dot) neighbourhood is shown as a red circle of 100 kpc radius.}
 \label{other tails}
\end{figure*}

In this study we find a very intriguing correspondence between particles generated by the 2-3 Gyr old merger that occurred at the M31 location and several MW dwarf galaxies, including the LMC. It suggests that the VPOS may take its origin from the M31 event. However, this conclusion has a major caveat: it currently only applies to about three-fourths of the VPOS dwarfs whose orbits lie below the MW disc plane (see Fig.~\ref{DwarfTracebackB}). The dwarfs that do not appear to be possibly associated with the tidal tail particles and currently lie above the MW plane are Crater II, Draco, Sculptor, and Ursa Minor. Of these, Sculptor's angular momentum is opposite of the other VPOS dwarfs (Fig. \ref{Angmom_pole}). In the future, we will study this set of VPOS dwarfs to verify whether or not they can also be related to a merger at the M31 location. The M31 major merger does produce two or even three tidal tails, which might be associated with one of the structures of the dwarf irregulars in the Local Group, identified by \citet{Pawlowski2013}. Figure \ref{other tails} shows two tidal tails from the M31 merger model 336. Stars (black) and gas (green) from one of the tails (which formed during an earlier passage of the merger) seem to approach the MW from the opposite side of the tidal tail we studied in this paper. Our future goal is to study whether the above dwarfs can be related to the same structure, which requires a complete study of both simulations and dwarf proper motions.

\section{Conclusion}
\label{sec: conclusion}
Using the \textit{Gaia} DR3 proper motions, we find that most of the VPOS dwarfs (including the Magellanic Clouds) are in their first approach towards the MW from a direction close to that of M31. For low-mass MW models, we find that the 3D position and 3D velocity of these dwarfs are consistent with that of the tidal tail particles generated by a recent major merger at the M31 location. For MW models with total masses lower than 5 $\times 10^{11} M_{\sun}$, we always find a small 6D difference between VPOS dwarfs and M31 tidal tail particles, with a statistical weight ranging from 0.35 to 1.2 $\sigma$. Accounting for the expected ram pressure effect due to the Galactic corona, we were able to associate twelve VPOS dwarfs with the M31 tidal tail, with another four of them coming from another direction. These four dwarfs might be associated with another M31 tidal tail, which would be the subject of another paper. This intriguing coincidence in the 6D phase space suggests that VPOS dwarf galaxies, including the LMC, could be TDGs. It could be the first clue of a material exchange between the M31 and the MW, as well as of the origin of the VPOS as a transient feature. 

\begin{acknowledgements}
We would like to thank the referee for many insightful comments and suggestions, which greatly strengthened the overall manuscript. We also thank Yongjun Jiao for a discussion about the MW potential models and for assistance with the galpy package. Istiak Akib would like to thank the Graduate Program in Astrophysics of the Paris Sciences et Lettres (PSL) University for funding this research. Marcel S. Pawlowski acknowledges funding via a Leibniz-Junior Research Group (project number J94/2020).
\end{acknowledgements}

\bibliographystyle{aa}
\bibliography{ref}

%\appendix
\begin{appendix}
\onecolumn

\section{Equations of the MW potential components}
\label{sec: MW Dark Matter Models}
\noindent The DM density for the Einasto profile used to describe the MW potential A to D is taken from \cite{Einasto1965} and is given by\\ 
$\rho(r) = \rho_0 \mathrm{ exp}\left[\left(-\dfrac{r}{r_0}\right)^n\right].$

\noindent The parameter values for this profile are taken from \cite{Jiao2021, Jiao2023}. The DM density for the NFW profile used in MW potential model E is taken from \cite{Navarro1997}, and is given by\\
$\rho(r) = \dfrac{\rho_0}{\left(\dfrac{r}{r_0}\right)\left(1+\dfrac{r}{r_0}\right)^2}.$ \\

\noindent The parameter values for this profile are taken from \cite{Eilers2019}. The CGM profile used in the MW potential model B is taken from the Model I of \cite{Wang2019}, and is given by\\ $\rho(r) = \rho_0 \left[ 1 + \left(\dfrac{r}{a}\right)^\alpha \right]^{\dfrac{\gamma - \beta} {\alpha}} \mathrm{exp}\left[ - \left(\dfrac{r}{r_{cut}}\right)^\zeta \right], $ \\ \\ \\
where $\rho_0 = 8.8764 \times 10^{-5}  \mathrm{ M_{\odot} pc^{-3}}, \alpha =2, \beta =2, \gamma = 0, \zeta =2, r_{cut} = 1300 \mathrm{kpc}$, and $a = 20 \mathrm{ kpc.} $

\section{Dwarf traceback in a massive MW (mass model E)}
\label{sec: Traceback MW E}

\begin{figure*}[h!]
\centering
    \includegraphics[width=\linewidth]{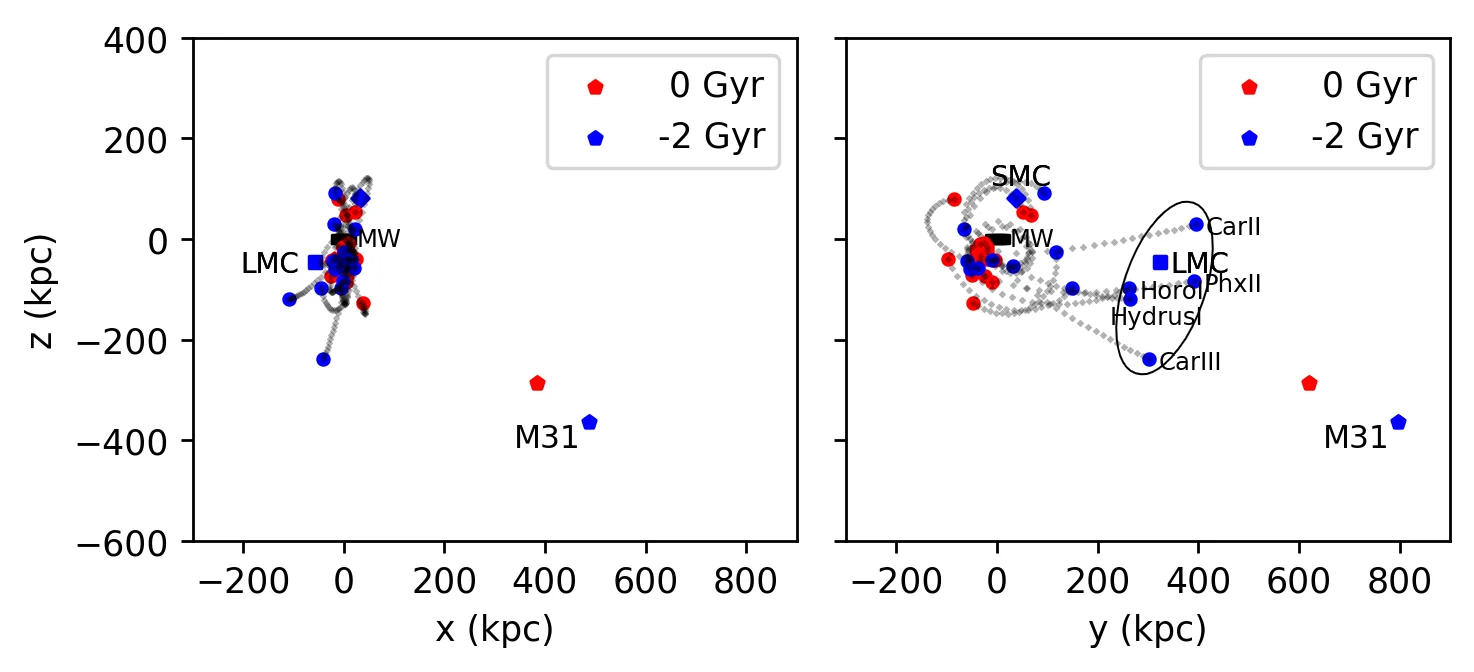}
 
 \caption[Dwarf Traceback]{Traceback of the VPOS dwarf galaxies with mean proper motion from \cite{Li2021} and \cite{Luri2021} under MW potential model E in a left-handed Galactocentric coordinate frame. The MW is the black line at (0,0,0). Red and blue colours correspond to the positions at the current epoch and at -2 Gyr, respectively. The black oval denotes the LMC-related dwarfs. }
 \label{Dwarf Traceback E}
\end{figure*}

\noindent Most dwarfs seem to be bound in high-mass MW models such as model E (figure \ref{Dwarf Traceback E}). Even the orbits of the high energy LMC dwarfs are significantly affected in terms of distance from the MW and infall direction. Hence, the tidal tail particles from the M31 merger models could not be associated with the LMC in the high-mass MW scenarios.

\section{Effect of the MW on the M31 merger }
\label{sec: MW on M31 merger}

\begin{figure*}[h!]
\centering
    \includegraphics[width=0.85\linewidth]{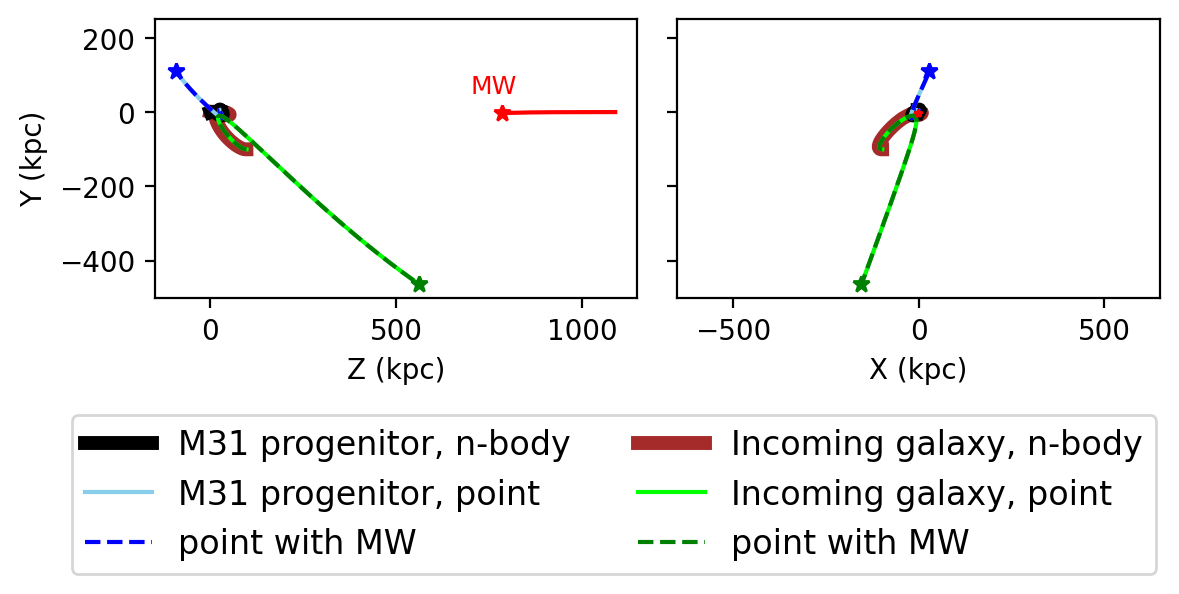}
 
 \caption[MW impact on M31 merger]{Point mass M31 main progenitor ($6.5\times10^{11} \mathrm{M_\odot}$) and the secondary interloper ($1.7\times10^{11} \mathrm{M_\odot}$) during the second passage (-5 Gyr to 0 Gyr) of M31 merger model 288 without (smooth line) and with (dashed line) the radially approaching MW point mass  ($8.14\times10^{11} \mathrm{M_\odot}$). Thick black and brown lines trace the orbit from the n-body simulations that result in the merger. The current Andromeda position is at (0,0,0), and the current MW is on the z-axis. Stars denote the positions at 0 Gyr.}
 \label{MW impact on M31 merger}
\end{figure*}

\noindent We considered whether the massive MW from model E could have impacted the orbital motions of the two galaxies that merged to form M31. In particular, the goal is to verify whether it might have affected the properties during the second passage that results in the tidal tail in consideration. A simple three-body simulation (M31 merger model 288 in Fig. \ref{MW impact on M31 merger}) shows that even the heaviest MW does not have a significant impact on the orbits of the main and secondary progenitor galaxies involved in the M31 merger. We find that the MW potential is negligible in the M31 neighbourhood and cannot affect the merger, as well as the formation of the tidal tail and associated dwarfs. However, one may note that the over-simplistic point mass simulations used in Fig.~\ref{MW impact on M31 merger} (thin lines and dashed lines) cannot explicitly reproduce the merger event, explaining why  galaxy orbits diverge, unlike the much more realistic N-body hydrodynamical simulations (thick lines).

\section{Effect of the M31 potential in the MW neighbourhood}
\label{sec: M31 Potential}

\begin{figure*}[h!]
\centering
    \includegraphics[width=\columnwidth]{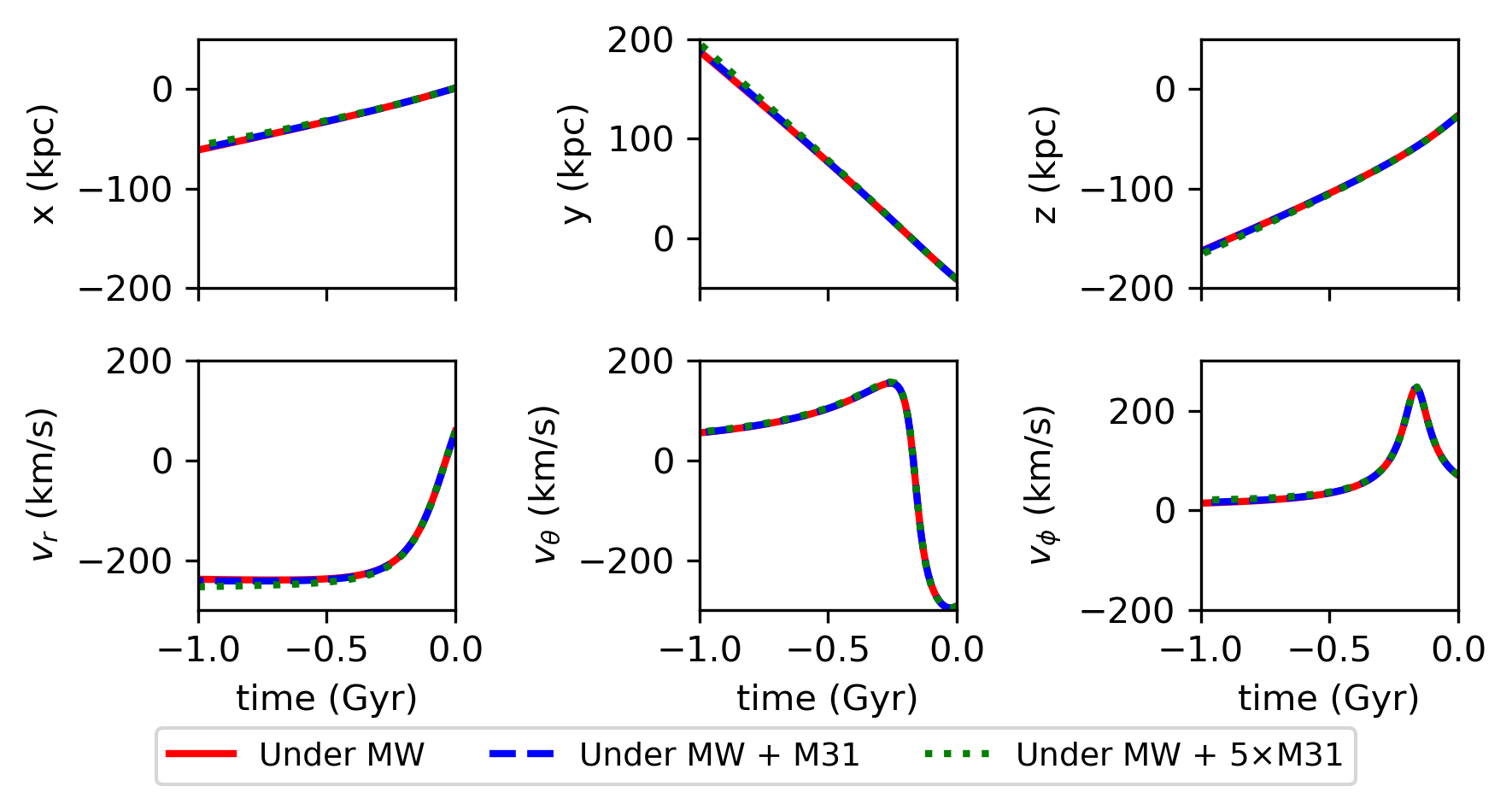}
\caption[Effect of M31 Potential]{Galactocentric coordinates of the LMC under MW model B (red), MW model B + M31 potential (blue dashed), and  MW model B + 5 $\times$ M31 potential (green dotted).}
\label{M31_effect}
\end{figure*}

\noindent In Fig. \ref{M31_effect} we show the galactocentric 6D coordinates of the LMC orbit under only a low mass MW (model B) and when M31 potential is added to it. The M31 potential, of mass $4.5\times10^{11} \mathrm{M_{\odot}}$ is taken from the model 371 of \cite{Hammer2025}, has been implemented using the moving object potential function within galpy for the M31 proper motion of $(\mu_{\alpha}, \mu_{\delta})$ = (75,0) \textmu as. The changes in the coordinates are very small, even when considering a 5 times more massive M31 (green dotted line) within our integration time and hence have been ignored.

\newpage

\section{LMC orbit correction due to the SMC collision, dynamical friction, and ram pressure}
\label{sec:LMC-SMC collison}

\begin{figure*}[h!]
\centering
 \includegraphics[width=\columnwidth]{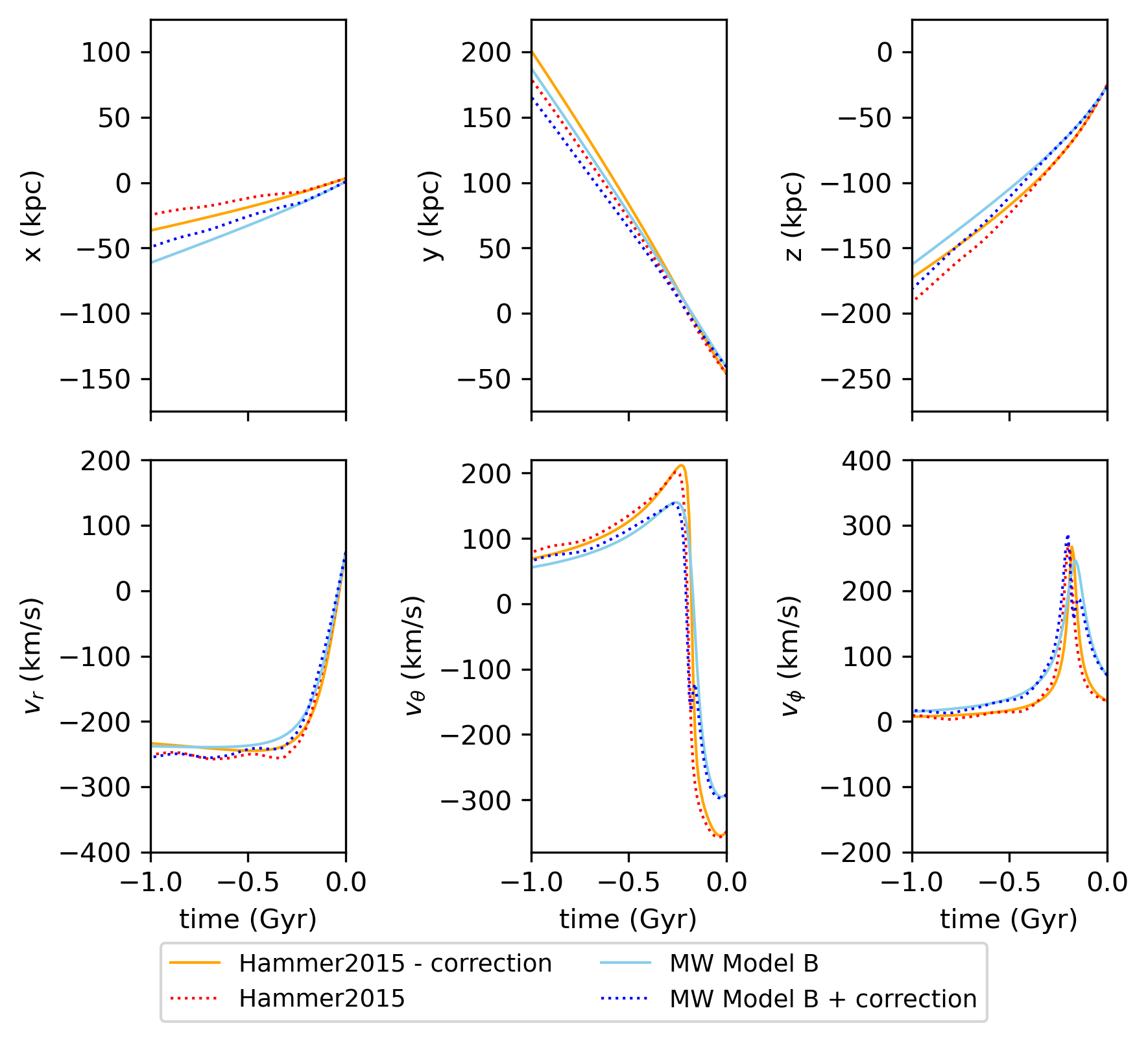}
 \caption[LMC SMC Encounter Correction]{Correction for the LMC-SMC encounter, dynamical friction, and ram pressure based on the hydrodynamic simulation done by \cite{Hammer2015} with a slightly different LMC proper motion and a MW mass model. Smooth lines are for the simple orbit calculations in \textit{galpy,} and the dotted lines indicate when the LMC-SMC encounter, dynamical friction, and ram pressure are taken into account.}
 \label{SMCcollision}
\end{figure*}

\noindent We used the LMC-SMC collision simulation (yellow in Fig. \ref{SMCcollision}) from \cite{Hammer2015}. sing their mass model ($6.9 \times 10^{11} M_{\odot}$) and their LMC position and velocity at the current epoch, we traced back the LMC in galpy (red). This traceback is only under the MW potential. The difference between these two orbits is considered to be the `correction,' which is due to the impact of the SMC, dynamical friction, and ram pressure on the LMC orbit. Our mass models and the LMC observations (distance and proper motion) are different from those used by \cite{Hammer2015}. So, we took the LMC traceback using our MW mass models and the LMC proper motion (blue in Fig. \ref{SMCcollision}, for MW model B) and added the correction over time to that. This approach assumes that the correction does not differ significantly with a small change in the observed LMC proper motion and the MW potential. A proper correction would require full hydrodynamical modelling of the collision with the new observational values and for each of the MW mass models.

\end{appendix}

\end{document}